\documentclass[12pt]{article}

\begin{document}

\baselineskip 20pt
\vspace*{-.6in}
\thispagestyle{empty}
\begin{flushright}
CALT-68-2322\\ CITUSC/01-006
\end{flushright}

\vspace{.5in} {\Large
\begin{center}
{\bf Comments on Born--Infeld Theory}\end{center}}
\bigskip
\begin{center}
 John H. Schwarz\\
\bigskip
\emph{California Institute of Technology}
\end{center}
\vspace{0.8in}

\begin{center}
\textbf{Abstract}
\end{center}
\begin{quotation}
\noindent The low-energy effective action of supersymmetric
D-brane systems consists of two terms, one of which is of the
Born--Infeld type and one of which is of the Chern--Simons type. I
briefly review the status of our understanding of these terms for
both the Abelian and non-Abelian cases.
\end{quotation}

\vspace{0.8in}

\centerline{\it Talk presented at Strings 2001}

\newpage

\pagenumbering{arabic}

\section{Single BPS D-Brane}

D-branes are surfaces on which open strings can end
\cite{Polchinski:1995mt} \cite{Polchinski:1996na}
\cite{Taylor:1997dy}. As such, their dynamics is described by {\it
open string field theory} -- e.g., of the type formulated by
Witten \cite{Witten:1986cc}. However, this can be difficult to
work with, so it is sometimes useful to consider the low energy
effective action obtained by integrating out all the massive
modes, keeping only the massless super-Maxwell multiplet. This is
too difficult, however. The best one can do is to keep terms in
which the massless fields are slowly varying at the string scale
-- keeping fields strengths, but not their derivatives. There is
no restriction on the size of field strengths relative to the
string scale, except that electric fields cannot exceed a certain
critical value.

The basic structure is always the sum of two terms
\begin{equation}
S = S_{DBI} + S_{CS}
\end{equation}
Here $S_{DBI}$ is the Dirac--Born--Infeld term (the only one in the case of
bosonic string theory), and $S_{CS}$ is the Chern--Simons term.
In the case of type II superstring theory, an $N$ D-brane system
is given by a $U(N)$ gauge theory \cite{Witten:1996im}.

Let us begin by considering a
single D-brane ($N=1$). In this case, ignoring fermi fields and
taking a flat 10d background, the action for a D9-brane is
\begin{equation}
S_{DBI} = T_9 \int d^{10} \sigma \sqrt{ - {\rm det}
(g_{\alpha\beta} + 2\pi \alpha' F_{\alpha\beta})}.
\end{equation}
Here $T_9$ is the D9-brane tension and $g_{\alpha\beta}$
is the pullback of the (flat) spacetime metric $\eta_{\mu\nu}$:
\begin{equation}
g_{\alpha\beta} = \eta_{\mu\nu} \partial_{\alpha} X^{\mu}
\partial_{\beta} X^{\nu},
\end{equation}
where $\sigma^{\alpha}$ ($\alpha = 0, 1, \ldots, p$) are the
world-volume coordinates, and $X^{\mu}(\sigma)$  ($\mu = 0, 1,
\dots, 9$) are the embedding functions.

The action $S_{DBI}$ has world-volume diffeomorphism invariance. A
natural gauge choice -- called {\it static gauge} -- is to
identify the first $p+1$ components of $X^{\mu}$ with
$\sigma^{\alpha}$. In this gauge the D9-brane action becomes
\begin{equation}
S_{DBI} = T_9 \int d^{10} \sigma \sqrt{ - {\rm det}
(\eta_{\alpha\beta} + 2\pi \alpha' F_{\alpha\beta})}.
\end{equation}
This formula was derived first for the bosonic string theory by
Fradkin and Tseytlin by computing the disk partition function
\cite{Fradkin:1985qd}. (For a review of Born--Infeld theory in the
context of string theory see \cite{Tseytlin:1999dj}.)

The corresponding actions for D$p$-branes with $p < 9$ can be
deduced by using T duality. The result, which agrees with dimensional
reduction, is
\begin{equation}
S_{DBI} = T_p \int d^{p+1} \sigma \sqrt{ - {\rm det} (\eta_{\alpha\beta} +
\partial_{\alpha} X^i \partial_{\beta} X^i + 2\pi \alpha' F_{\alpha\beta})}.
\end{equation}
The index $i= p+1, \ldots, 9$ labels the $9-p$ directions
transverse to the Dp-brane. The world-volume scalars $X^i$ can be
regarded as Goldstone bosons associated to broken translational
symmetries.

It is remarkable that the entire $\alpha'$ expansion of the low
energy effective action of open strings -- for slowly varying
fields -- can be encoded in such simple formulas. In particular,
as observed by Bachas \cite{Bachas:1996kx}, for a D0-brane this gives
\begin{equation}
T_0 \int d \sigma^0 \sqrt{ 1 - \partial_0 X^i \partial_0 X^i},
\end{equation}
which is the standard action for a relativistic particle of mass $T_0$.

It is natural to wonder whether something analogous might be
possible for an effective low energy theory of closed strings in
terms of the gravity supermultiplet. No such formula is known, but
if one could be found, it would be very interesting. It could be
useful for exploring whether curvatures are bounded and whether
some spacetime singularities are thereby evaded.

\subsection{Supersymmetrization}

The supersymmetrization of the D-brane action was worked out  in 1996
by several different groups \cite{Aganagic:1997pe}
\cite{Cederwall:1997ri} \cite{Bergshoeff:1997tu}.
The idea is to embed the
D-brane in superspace $(X^{\mu}, \theta_1^a, \theta_2^a)$, where
$(\theta_1, \theta_2)$ are MW spinors.
The global ${\cal N} = 2$, $D=10$ supersymmetry is realized on
superspace in the usual way ($ \delta \theta = \epsilon$, $\delta
X^{\mu} = \bar \epsilon \Gamma^{\mu}\theta$). The D-brane action
is constructed out of the supersymmetry invariants
\begin{equation}
\Pi^{\mu}_{\alpha} = \partial_{\alpha} X^{\mu} - \bar \theta\Gamma^{\mu}
\partial_{\alpha} \theta
\end{equation}
and $\partial_{\alpha}\theta$. The $\theta$'s would give twice the
number of desired fermions (16 instead of 8) except for the fact
that half of them are compensated by a local fermionic symmetry:
called {\it kappa symmetry}.

The addition of the $\theta$'s plus the requirements of global
supersymmetry and local kappa symmetry determine the action. One
finds
\begin{equation}
S_{DBI} = T_p \int d^{p+1} \sigma \sqrt{ - {\rm det} ( G_{\alpha\beta}
+ 2\pi \alpha' {\cal F}_{\alpha\beta})}
\end{equation}
\begin{equation}
S_{CS} = \pm T_p \int \Omega_{p+1} ,
\end{equation}
where
\begin{equation}
G_{\alpha\beta} = \eta_{\mu\nu} \Pi_{\alpha}^{\mu} \Pi_{\beta}^{\nu}
\end{equation}
\begin{equation}
{\cal F}_{\alpha\beta} = F_{\alpha\beta} - B_{\alpha\beta} - b_{\alpha\beta}.
\end{equation}
Here $B$ is the pullback of NS-NS 2-form background field and $b$
is a two-form involving the fermi fields, which in the IIA case is
\begin{equation}
b = - \bar\theta \Gamma_{11} \Gamma_{\mu} d\theta (dX^{\mu} + {1\over 2}
\bar\theta \Gamma^{\mu} d\theta).
\end{equation}

The familiar result for $\Omega_{p+1}$ (due to Douglas
\cite{Douglas:1995bn}) in the presence of R-R background fields is
\begin{equation}
\Omega_{p+1} = \left(C e^{2\pi\alpha'F}\right)_{p+1} ,
\end{equation}
where $C = \sum C^{(n)}$ is a formal sum of R-R $n$-form fields.
($n$ is odd for IIA and even for IIB.) For $C$ constant this is
closed, but there is an additional piece involving the $\theta$'s
that contributes to
\begin{equation}
I_{p+2} = d \Omega_{p+1}.
\end{equation}
It has the structure
\begin{equation}
\left\{e^{2 \pi \alpha'\cal F} f(\Pi^{\mu},
d\theta)\right\}_{p+2}.
\end{equation}

\subsection{Static Gauge}

We will consider the gauge-fixed super D-brane action for the $p =
9$ case only. The formulas for $p < 9$ can be inferred by
dimensional reduction. As before, the local diffeomorphism
symmetry is used to identify the embedding functions $X^{\mu}$
with the world volume coordinates $\sigma^{\alpha}$. In addition,
the local kappa symmetry is used to eliminate half of the $\theta$
coordinates. A simple choice that preserves the manifest 10d
covariance is to simply set one of the two $\theta$'s, $\theta_2$
say, to zero \cite{Aganagic:1997nn}. This has the remarkable
consequence of completely eliminating the Chern--Simons term.

Renaming $\theta_1 = \lambda$ and setting
$2\pi\alpha' =1$ leaves the action
\begin{equation}
\int d^{10}\sigma \sqrt{- {\rm det} ( \eta_{\alpha\beta} +
F_{\alpha\beta} - 2 \bar \lambda \Gamma_{\alpha} \partial_{\beta}
\lambda + \bar\lambda \Gamma^{\rho}\partial_{\alpha}\lambda
\bar\lambda \Gamma_{\rho}\partial_{\beta}\lambda)}.
\end{equation}
This is the ${\cal N} = 1$, $D=10$ super-Maxwell theory
supplemented by higher-dimension interaction terms. The latter are
very special, because in addition to the 16 linearly realized
supersymmetries of the free theory there are 16 additional
nonlinearly realized supersymmetries. That is why this formula is
reminiscent of the Volkov-Akulov action \cite{Volkov:1973ix} --
$\lambda$ can be interpreted as the Goldstone field for the broken
supersymmetries. The formula is unique up to the freedom of field
redefinitions. It would have been extremely difficult to discover
if one had specialized to the static gauge before
supersymmetrizing the action.

When this D9-brane action is dimensionally reduced to give the
D$p$-brane action:

\noindent $\bullet$ 16 supersymmetries and $p+1$ translation
symmetries are linearly realized.

\noindent $\bullet$ 16 supersymmetries and $9-p$ translation
symmetries are nonlinearly realized and correspond to Goldstone
modes on the world volume.

It would be very interesting to rederive this formula by computing
the superstring disk partition function in the presence of the
appropriate boundary interactions. This ought to be a tractable
extension of the calculations described in recent papers
\cite{Kraus:2000nj} \cite{Takayanagi:2000rz}.

\section{Non-Abelian Generalizations}

When one has $N$ coincident type II D$p$-branes the world-volume
theory is a $U(N)$ gauge theory. As such, it should be given by a
non-Abelian generalization of the formulas of the preceding
section. The explicit construction of such an action is a
difficult problem that has been studied extensively
(starting with \cite{Argyres:1990qr}), but is not
yet completely settled.

Tseytlin proposed a specific recipe for generalizing
Abelian formulas to non-Abelian ones \cite{Tseytlin:1997cs}.
His proposal -- referred to
as the {\it symmetrized trace prescription} -- works as follows.  An
expression in the Abelian theory, such as $\sqrt{-{\rm det} (
\eta_{\alpha\beta} + F_{\alpha\beta})},$ has an expansion of the
form
\begin{equation}
1+ {1\over 4} F^2 -{1 \over 8} ( F^4 - (F^2)^2) + \dots,
\end{equation}
where $F^2 = F_{\alpha\beta} F^{\beta\alpha} $, etc. In the
non-Abelian case, $F$ is also a hermitian $N\times N$ matrix.
Tseytlin's proposal is to take the trace of each term in the
expansion, and to resolve the ordering ambiguities by averaging
over all possible choices. The result is denoted
\begin{equation}
 {\rm Str} \sqrt{-{\rm det} ( \eta_{\alpha\beta} + F_{\alpha\beta})}.
\end{equation}
Studies by Hashimoto and Taylor \cite{Hashimoto:1997gm} and others
suggest that this is a correct rule through terms of order $F^4$,
but that it fails at higher orders.

\subsection{The Chern--Simons Term}

It has been clear since the work of Douglas \cite{Douglas:1995bn}
that D-branes can carry R-R charges associated with lower
dimensional branes. More recently, it has been realized that in
the non-Abelian case they can also carry charges associated with
higher dimension D-branes \cite{Taylor:2000pr}
\cite{Myers:1999ps}.

Myers discovered an interesting part of the the answer by
exploring consistency with T duality  \cite{Myers:1999ps}.
He focused on the dependence
on the bosonic fields $A_{\alpha}$ and $X^i$, each of which are
now $N \times N$ matrices in the static gauge with all fermi
fields set to zero. He included the dependence on $B$ and $C$
background fields. For the Chern--Simons term he obtained the
result
\begin{equation}
S_{CS} = T_p \int {\rm Str} \left( P [ e^{i I_X I_X} C e^B] e^F \right) .
\end{equation}

This is a subtle formula that requires some explanation. First of
all, $C = \sum C^{(n)}$, as before.  $P[\ldots]$ means the
pullback to the world volume, since $B$ and $C$ are bulk fields.
$X$ refers to the $9-p$ scalars $X^i$, which are now $N\times N$
matrices. The operation $I_X I_X$ acting on an $n$-form gives an
$n-2$-form. For example,
\begin{equation}
I_X I_X C^{(2)} = X^j X^i C^{(2)}_{ij} =
{1 \over 2} C^{(2)}_{ij} [X^j , X^i]
\end{equation}
Moreover, in the pullback of a function $f(x^{\alpha}, x^i)$, the
matrices $X^i$ need to substituted for the bulk coordinates $x^i$.
This requires an ordering prescription, since $[X^i, X^j] \neq 0$.
The proposed formula is
\begin{equation}
P[f] = {\rm exp}(X^i {\partial \over \partial x^i})
f(\sigma^{\alpha}, x^i) |_{x^i = 0}.
\end{equation}

These rules are sufficiently subtle that it is extremely hard to
check whether or not $S_{CS}$ is invariant under gauge
transformations of the type $C \to C + d \Lambda$. I would guess
that this fails at some point, but I am not sure. It does work at
low orders.

A crucial feature of this formula, for which there is a lot of
evidence by now, is that multi D-brane systems can be sources of
higher D-brane charge as well as lower D-brane charge, since all
the R-R fields appear. This is to be contrasted with the Abelian
case where $(Ce^F)_{p+1}$ only depends on $C^{(p+1)}, C^{(p-1)},
\dots$.

Myers discovered a {\it dielectric effect} in which an R-R field
strength can cause the brane to expand into new dimensions. For
example, a system of $N$ D0-branes in the presence of an electric
$F^{(4)} = d C^{(3)}$ becomes a fuzzy two-sphere with $[X^i, X^j]
\sim N \epsilon^{ijk} X^k$. (See \cite{Taylor:1999gq} for earlier
related work.) For large $N$ this describes an ordinary $S^2$ with
radius proportional to $N$. This can be interpreted as a spherical
D2-brane with $N$ D0-branes bound to it. The Myers effect is
relevant to the appearance of ``giant gravitons'' on the AdS side
of the AdS/CFT correspondence \cite{McGreevy:2000cw}.

\subsection{Supersymmetrization}

Part of the rationale for Tseytlin's symmetrized trace
prescription is that a field strength commutator $[F_{ij},
F_{kl}]$ is proportional to $[D_i, D_j] F_{kl}$ and thus can be
regarded as being higher-order in derivatives. This reflects an
inherent ambiguity in the meaning of ``slowly varying fields'' in
the non-Abelian case. This might be resolved by requiring that the
action have all the desired symmetries: supersymmetry, kappa
symmetry, etc.

In the case of $N$ coincident D-branes the supersymmetric $U(N)$
world-volume theory should again have as its physical field
content gauge fields $A_{\alpha}$, transverse scalars $X^i$, and
fermi fields $\lambda$ -- this time all in the adjoint of the
$U(N)$ Lie algebra. An interesting question is how to generalize
the $U(1)$ formulation with local diffeomorphism invariance and
local kappa symmetry to $N>1$. It is natural to suppose that one
should start with world volume fields $A_{\alpha}(\sigma)$,
$X^{\mu}(\sigma)$, $\theta(\sigma)$, each of which is $U(N)$
valued. Then, in order to end up with the right physical degrees
of freedom, one would need $U(N)$ generalizations of the
diffeomorphism and kappa symmetries. These would allow us to
choose a gauge  that would restrict $X^{\mu} \to X^i$  and $\theta
\to \lambda$.

The case of D9-branes is somewhat special in that there are no
transverse directions $X^i$. Thus, in static gauge, the only
world-volume fields are the gauge fields $A_{\alpha}$ and the
fermi fields $\theta$. Still this is quite general, because the
results for $p < 9$ can be deduced by dimensional reduction.  One
still needs kappa transformations in the adjoint of $U(N)$ so that
a gauge choice can reduce $\theta$ to $\lambda$.

This kind of a set-up has been explored recently by Bergshoeff, de
Roo, and Sevrin \cite{Bergshoeff:2000ik}. They carried out an
iterative analysis that allowed them to deduce the action up to a
certain order. Specifically, they determined terms in $S_{CS}$
with the structures $\theta D \theta$, $\theta D \theta F$, and
$\theta D \theta F^2$, and in $S_{DBI}$ with the structures $1$,
$F^2$, $\theta D \theta$, $\theta D \theta F$, and $\theta D
\theta F^2$. Up to this order they succeeded in obtaining unique
results with all the desired properties. They also gave the
formulas in the gauge $\theta_2  =0$, $\theta_1 = \lambda$. As in
the Abelian case, $S_{CS}$ does not contribute in this gauge. They
observed, in particular, that the $\bar\lambda D \lambda F^2$
terms cannot be be expressed in terms of symmetrized traces.

The iterative analysis of Bergshoeff, de Roo, and Sevrin is
technically difficult and cannot be pushed much further. It seems
to me that the best hope for complete results, generalizing those
of the Abelian case, would use an approach that does not invoke
the static gauge at the outset. This would seem to require a
matrix generalization of diffeomorphism invariance, but I doubt
that such a thing is possible.

\section{Conclusion}

In conclusion, we have presented a review of the structure of low
energy effective actions for D-branes. We saw that D-brane world
volume actions are always given as the sum of a
Dirac--Born--Infeld term and a Chern--Simons term and each term
contains a lot of important information.

It would be desirable to have explicit exact results for the
non-Abelian case so that one could explore the non-Abelian
generalization of various effects that have been studied in the
Abelian case. These include classical solutions that describe
various sorts of solitons and brane configurations, as well as
physical effects associated with electric fields approaching
limiting values.

A powerful approach that has received a great deal of attention
lately is BSFT: boundary string field theory or background
independent string field theory \cite{Witten:1992qy}. This
provides the logical basis for deriving D-brane effective actions
in terms of disk partition functions with appropriate boundary
interactions. The BSFT approach allows one to formulate unstable
D-brane systems and to test some of Sen's conjectures regarding
tachyon condensation \cite{Sen:1998sm}.  This has been done with
notable success in recent works \cite{Gerasimov:2000zp}
\cite{Kutasov:2000qp}. It also provides another approach to to
studying the formation of various sorts of solitons and to
formulating the non-Abelian Born--Infeld problem. Unfortunately,
the relevant path integrals may not be amenable to analytic
evaluation.

\section*{Acknowledgments}
This work was supported in part by the U.S. Dept. of Energy under
Grant No. DE-FG03-92-ER40701.

\end{document}